 \documentclass[final,5p,times,twocolumn,authoryear,article]{elsarticle}

\usepackage{amssymb}
\usepackage{lipsum}
\usepackage{multicol}
\usepackage{float}
\journal{SLAS Technology}

\begin{document}

\begin{frontmatter}



\title{Accelerating drug discovery with Artificial: a whole-lab orchestration and scheduling system for self-driving labs}


\author[first]{Yao Fehlis, Paul Mandel, Charles Crain, Betty Liu, David Fuller}
\affiliation[first]{organization={Artificial Inc.}}  % Removed the comma

\begin{abstract}
Self-driving labs are transforming drug discovery by enabling automated, AI-guided experimentation, but they face challenges in orchestrating complex workflows, integrating diverse instruments and AI models, and managing data efficiently. Artificial addresses these issues with a comprehensive orchestration and scheduling system that unifies lab operations, automates workflows, and integrates AI-driven decision-making. By incorporating AI/ML models like NVIDIA BioNeMo—which facilitates molecular interaction prediction and biomolecular analysis—Artificial enhances drug discovery and accelerates data-driven research. Through real-time coordination of instruments, robots, and personnel, the platform streamlines experiments, enhances reproducibility, and advances drug discovery.
\end{abstract}




\end{frontmatter}

%\tableofcontents



\section{Introduction}
\label{introduction}

The landscape of drug discovery has long been characterized by a multitude of challenges, including the high costs of research and development, lengthy timelines, and a significant rate of failure during clinical trials ~\citep{blanco2023role,udegbe2024machine,khanna2012drug,moffat2017opportunities}. These hurdles have necessitated the exploration of innovative approaches that can streamline drug development processes. Traditional methods are often labor-intensive and prone to human error, which can lead to inefficiencies and inconsistencies in result fidelity. As a response to these challenges, self-driving laboratories have emerged as a promising solution, leveraging automation and artificial intelligence (AI) to enhance the efficiency and efficacy of experimental workflows.

Self-driving labs are designed to automate various aspects of the drug discovery process, from hypothesis generation to experimental execution and data analysis~\citep{edfeldt2024data,rapp2024self}. These labs integrate robotic systems that can autonomously perform experiments based on pre-defined protocols, significantly reducing human involvement and minimizing the chances for error. This automation is not just about increasing throughput but also about maintaining a high degree of precision and reproducibility, which are critical factors in drug development. As self-driving labs continue to evolve, they are being increasingly complemented by advances in artificial intelligence, which plays a pivotal role in orchestrating various laboratory activities and optimizing resource allocation.

A crucial aspect of modern drug discovery is the use of virtual screening~\citep{lavecchia2013virtual,zhou2024artificial,zhu2022comprehensive,murugan2022artificial}, molecular simulations~\citep{adelusi2022molecular,casalini2021not,shukla2021molecular,adediwura2024understanding}, and AI-driven computational methods~\citep{mak2024artificial, deng2022artificial,Dodero-Rojas2024.07.17.603864}. Virtual screening allows researchers to rapidly evaluate large libraries of chemical compounds, prioritizing those most likely to exhibit therapeutic potential. Molecular simulations provide insights into drug-target interactions, protein folding, and binding affinities, significantly reducing the need for exhaustive trial-and-error experimentation. AI further enhances these processes by applying machine learning algorithms to predict molecular properties, optimize lead compounds, and refine drug candidates through iterative learning. By integrating these computational tools into experimental workflows, researchers can accelerate drug discovery, reduce costs, and improve the likelihood of identifying effective therapeutics. The synergy between in silico methodologies and automated laboratory processes represents a transformative shift in how drugs are developed, enabling more efficient and data-driven decision-making.

Artificial intelligence in the context of drug discovery offers powerful capabilities such as predictive modeling, machine learning-based data analysis, and the automation of repetitive tasks~\citep{schauperl2022ai,marco2024augmenting,obaido2024supervised,colliandre2023bayesian,pyzer2018bayesian,guo2024artificial}. These AI-driven technologies can process vast amounts of data generated from experiments, facilitating more informed decision-making and accelerating the pace of drug discovery~\citep{zhu2020big,glicksberg2019leveraging,vergetis2021assessing}. Furthermore, the integration of AI with laboratory automation establishes a framework in which complex experimental designs can be implemented with greater accuracy and less manual oversight~\citep{abolhasani2023rise,delgado2023research,sanders2023biological}. By enabling intelligent decision-making and optimizing workflows, AI enhances the efficiency of self-driving labs and contributes to more rapid identification of promising drug candidates.

However, the effective implementation of AI in drug discovery is not without challenges. A predominant issue that hampers AI performance is the existence of data silos—discrete sets of data that remain isolated and uncoupled from broader datasets~\citep{denton2021data, wibowo2017machine, patel2019overcoming, kasturi2014interconnectivity, ziegler2021metadata}. The potential for AI optimization hinges on the availability of comprehensive and high-quality datasets; thus, when relevant data is dispersed across various studies or proprietary platforms, researchers may find it difficult to train robust AI models. Addressing this issue requires strategic initiatives aimed at fostering data sharing and creating standardized data formats that facilitate integration. As the drug discovery landscape increasingly embraces AI, collaboration among institutions and industries becomes paramount to democratize access to data, thereby unleashing the full potential of AI-driven methodologies~\citep{cheng2020importance,khanna2012drug,peeva2025unlocking}.

In addition to data fragmentation, integrating AI models into scientific labs presents further challenges due to the diverse and noisy nature of experimental data~\citep{gadiya2023fair,blatter2022big,alharbi2022selection,alharbi2023fair}. AI models must be tailored to specific scientific needs, ensuring interpretability while seamlessly interacting with laboratory systems. Successful AI deployment requires robust computational pipelines capable of handling complex workflows, standardizing data preprocessing, and maintaining reproducibility. Without well-structured data pipelines, AI-driven insights may suffer from inconsistencies, ultimately reducing their reliability in guiding drug discovery efforts. Therefore, a well-orchestrated integration framework is necessary to ensure that AI not only enhances experimental workflows but also generates reproducible and actionable outcomes.

To address these challenges and maximize the potential of AI-driven self-driving laboratories, this paper presents Artificial, a platform designed to orchestrate laboratory workflows, integrate AI-driven decision-making, and facilitate seamless data management. Artificial enables the incorporation of AI models, such as those implemented in NVIDIA BioNeMo NIMs (containerized, pre-trained AI models with APIs), into self-driving virtual screening and experimental processes. This integration enhances the efficiency of drug discovery by enabling automated, data-driven decision-making and improving the reproducibility of experimental workflows.

In this paper, we present a proof-of-concept case study demonstrating the orchestration of NVIDIA BioNeMo models within a self-driving virtual screening workflow in a dry lab setting. This study highlights how Artificial automates AI model deployment, optimizes computational resource allocation, and facilitates seamless data exchange, ultimately accelerating the identification of promising drug candidates. While the case study focuses on dry lab applications, the underlying methodology and benefits of Artificial extend to both dry and wet lab environments, enabling more efficient execution of AI-driven drug discovery across the entire R\&D pipeline.

We further discuss how Artificial enhances laboratory automation by integrating AI models into structured workflows, optimizing resource utilization, and ensuring reproducible data-driven experimentation. The platform’s orchestration engine automates workflow planning, scheduling, and data consolidation, significantly improving efficiency in iterative experimentation. By streamlining AI-driven analyses and facilitating rapid hypothesis testing, Artificial enables self-driving labs to shorten discovery timelines and improve the success rate of drug development.

\section{Artificial: a whole-lab orchestration and scheduling system}
\label{artificial}

\begin{figure*}[ht]
    \centering
    \includegraphics[width=0.90\textwidth]{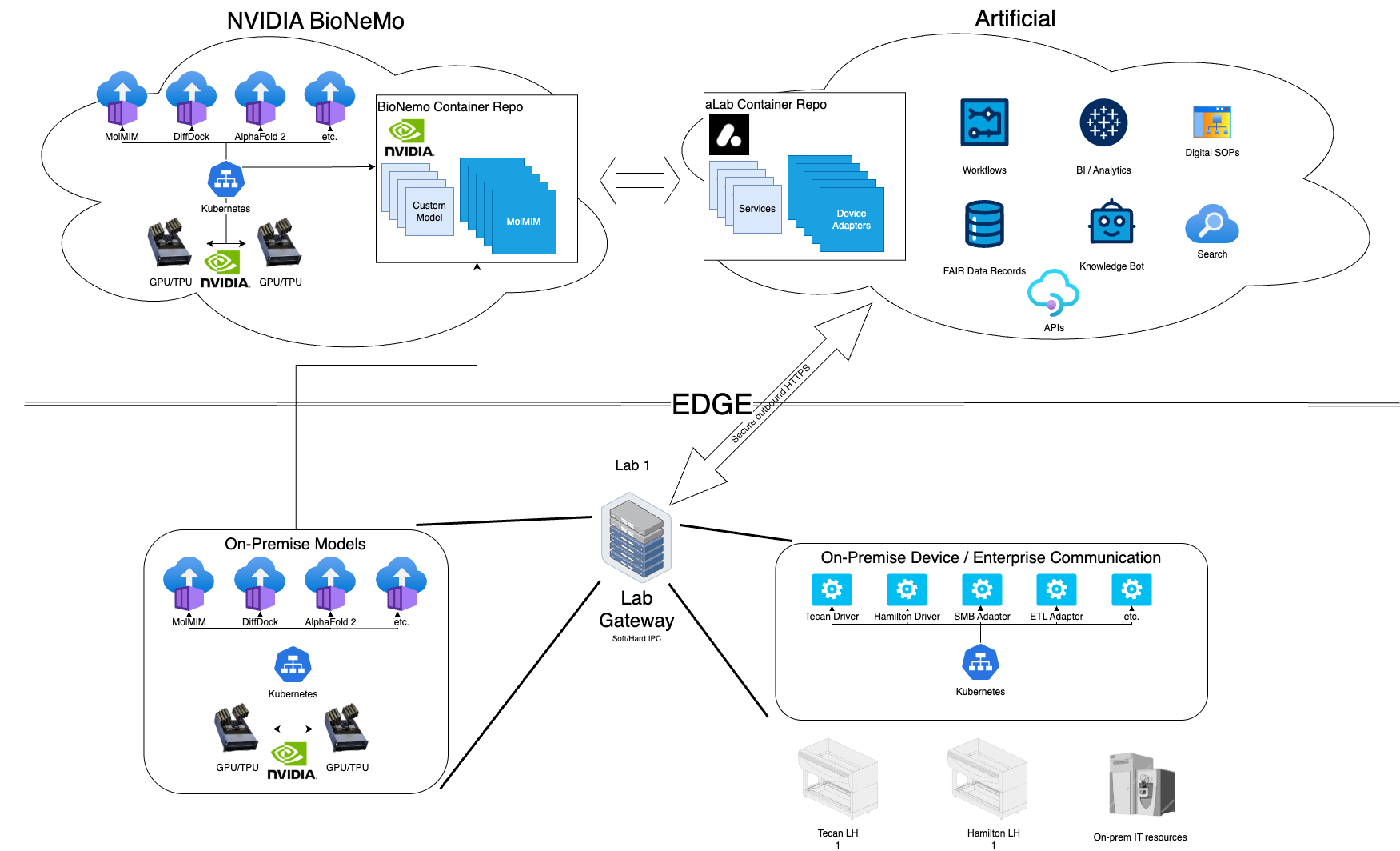}  % Reduce width slightly
    \caption{The Artificial stack.}
    \label{fig:stack}
\end{figure*}

Figure \ref{fig:stack} illustrates a modular and scalable Artificial Orchestration Platform designed to streamline laboratory workflows, automate processes, and connect people, samples, robots, and instruments. It integrates data, automation, and informatics through a series of interconnected components and services. Below is a detailed explanation of its structure and logic: \\
\\

\textbf{Core Components} \\
\begin{enumerate}
 \item \textbf{Web Apps (User-Facing Interfaces)}:
 \begin{itemize}
  \item Accessible on any device, these tools are designed for scientists and operators to interact with and control the lab ecosystem:
  \begin{itemize}
      \item \textbf{Labs}: Enables building and managing a digital twin of the lab environment, mapping its operations, instruments, and workflows.
      \item \textbf{Assistants}: Provides interactive guides and instructions for scientists and lab operators to follow manual or semi-automated processes.
      \item \textbf{Workflows}: Facilitates the definition, configuration, and management of R\&D processes, supporting repeatability and process optimization.
      \item \textbf{LabOps}: Acts as a central hub for monitoring, running, and orchestrating both manual and automated R\&D workflows.
      \item \textbf{Digital Twin}: A 3D visualization layer that mirrors the physical lab's environment, allowing simulation, monitoring, and optimization of experiments and processes.
  \end{itemize}
 \end{itemize}

 \item \textbf{Services}:
 \begin{itemize}
     \item These backend components provide the computational power to manage orchestration, scheduling, data storage, and optimization:
     \begin{itemize}
         \item \textbf{Orchestration}: Handles planning and request management for lab operations using a simplified dialect of Python, or a graphical interface.
         \item \textbf{Scheduler/Executor}: An orchestration engine that uses heuristics, constraints, and batching to ensure efficient resource allocation and workflow execution.
         \item \textbf{Data Records}: Consolidates lab data, including results and logs, into an accessible and organized repository.
     \end{itemize}
 \end{itemize}

 \item \textbf{Lab API (Connectivity Layer)}:
 \begin{itemize}
     \item A central interface supporting \textbf{GraphQL}, \textbf{gRPC}, and \textbf{REST} protocols, enabling integration between lab hardware, software, and external systems.
     \item API endpoints provide access to:
     \begin{itemize}
         \item \textbf{Lab States and Events}: Real-time monitoring of instrument status, process progress, and more.
         \item \textbf{Client Libraries}: Pre-built libraries for developers to interact with lab data and automation.
     \end{itemize}
 \end{itemize}

 \item \textbf{Adapters and Communication Protocols}:
  \begin{itemize}
     \item Enables \textit{secure, local, and global} communication between instruments, schedulers, informatics systems, and GPUs/TPUs via:
     \begin{itemize}
         \item \textbf{HTTPS}
         \item \textbf{gRPC (including SiLA)}
         \item \textbf{Local APIs (including SciKit, TensorFlow/Keras, and PyTorch for AI/ML)}
     \end{itemize}
     \item Supports the full Python programming language with its rich ecosystem. 
 \end{itemize}

 \item \textbf{Your Informatics (Integration with Lab Systems)}:
 \begin{itemize}
     \item The platform connects to standard informatics tools like:
     \begin{itemize}
         \item \textbf{LIMS (Laboratory Information Management System)}: Inventory and sample management.
         \item \textbf{ELN (Electronic Lab Notebook)}: Data management and collaboration.
         \item \textbf{On-premises or cloud IT infrastructure}: Integration with data lakes, business intelligence systems, etc.
     \end{itemize}
 \end{itemize}

 \item \textbf{Your Automation}:
 \begin{itemize}
     \item This layer interfaces with local hardware and software schedulers (e.g., Cellario, Momentum) and \textbf{instrument drivers} (e.g., Tecan, Thermo, Hamilton, Agilent).
     \item Ensures secure and efficient connection through \textbf{REST APIs} with authentication.
 \end{itemize}
\end{enumerate}

\textbf{Flow and Logic} \\
\begin{enumerate}
    \item \textbf{Data Flow}:
    \begin{itemize}
        \item Samples, robots, and instruments feed data into the system via the \textbf{Lab API}, which consolidates and processes it using backend services such as \textbf{Orchestration} and \textbf{Data Records}.
        \item The \textbf{Digital Twin} visualizes lab activities in real-time, providing an overview of workflows and system health.
    \end{itemize}
    
    \item \textbf{Automation and Optimization}:
    \begin{itemize}
        \item Automated workflows are scheduled and executed using the \textbf{Scheduler/Executor and Orchestration} components.
        \item The system ensures efficient resource usage by considering constraints and heuristics.
    \end{itemize}

    \item \textbf{Integration with Existing Systems}:
    \begin{itemize}
        \item The platform bridges legacy informatics systems (LIMS, ELN) with modern cloud and on-premise infrastructure, ensuring seamless operation across the lab.
    \end{itemize}

    \item \textbf{User Interaction}:
    \begin{itemize}
        \item Scientists interact with the platform through \textbf{Web Apps}, allowing them to define workflows, monitor progress, and manage lab operations with minimal complexity.
    \end{itemize}

    \item \textbf{Scalability and Security}:
    \begin{itemize}
        \item Built for deployment on both \textbf{cloud (EKS/AKS)} and \textbf{local environments (MicroK8s)}, ensuring flexibility and scalability.
        \item Communication is secured with \textbf{SSL encryption} to protect sensitive lab data.
    \end{itemize}
\end{enumerate}

\section{A case study}

\begin{figure}[H]
    \centering
    \includegraphics[scale=0.25]{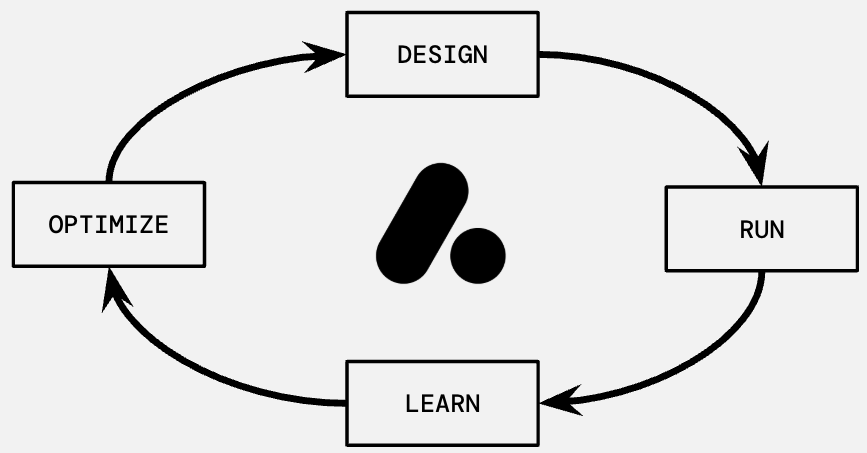}
    \caption{The self-driving cycle.}
    \label{fig:cycle}
\end{figure}

We would like to present a case study on how Artificial powers self-driving labs by integrating design, optimization, execution, and AI-driven learning into an automated workflow.

Figure  \ref{fig:cycle} illustrates how the Artificial platform supports self-driving labs by enabling seamless integration of AI into laboratory workflows. The process is divided into four interconnected phases: Design, Run, and Learn, Optimize, creating a continuous improvement cycle powered by AI and automation:

\begin{enumerate}
\item \textbf{Design}:\\
The platform provides a Digital Lab Toolbox to design end-to-end workflows, integrating instruments, manual steps, labware, and AI systems. This ensures streamlined and reproducible processes tailored to specific experiments.

\item \textbf{Run}:\\
LabOps enables real-time monitoring, tracking, and execution of experiments from anywhere. This ensures flawless execution, eliminates errors, and keeps experiments on schedule by automating task coordination.

\item \textbf{Optimize}: \\
The LabOps engine optimizes lab performance by intelligently coordinating tasks and batch processes. It considers real-time availability of instruments, personnel, and resources to enhance efficiency and reduce downtime.

\item \textbf{Learn}: \\
The platform logs all scientific and process data, providing a complete 360° context of each experiment. This data is leveraged by AI to generate insights, test hypotheses, and drive continuous improvements in workflows.
This closed-loop system empowers self-driving labs to optimize operations, accelerate discovery, and achieve reproducibility through AI-driven automation and learning.
\end{enumerate}

This closed-loop system empowers self-driving labs to optimize operations, accelerate discovery, and achieve reproducibility through AI-driven automation and learning.

\section{A case study: orchestrating NVIDIA BioNeMo models in self-driving virtual screening}

Virtual screening is crucial in drug discovery as it enables the rapid and cost-effective identification of potential drug candidates from vast compound libraries, significantly reducing the time and resources needed for experimental testing. It helps prioritize the most promising compounds for further development and testing.

NVIDIA BioNeMo is a platform for accelerating AI model development and deployment for digital biology applications~\citep{john2024bionemo}. Designed to accelerate drug discovery, protein engineering, and virtual screening processes, BioNeMo provides pre-trained AI models as containerized microservices (NIM, i.e. NVIDIA Inference Microservices) and tools to predict molecular interactions, facilitate protein folding, and analyze biomolecular data with a high degree of acceleration and efficiency~\citep{philippidis2024gtc}.

\begin{figure*}[h]
    \centering
    \includegraphics[width=0.95\textwidth]{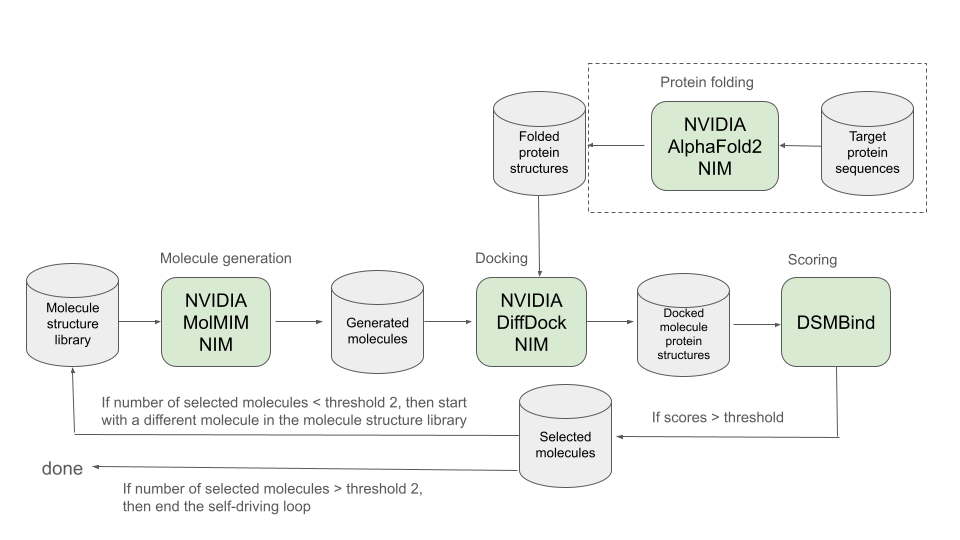}  % Reduce width slightly
    \caption{A proof of concept use case for self-driving virtual screening with NVIDIA BioNeMo models.}
    \label{fig:poc}
\end{figure*}

We have demonstrated the capabilities of workflows and model integration in Artificial using a proof of concept (PoC) use case for self-driving generative virtual screening. This PoC effectively showcases how NVIDIA's BioNeMo NIM microservices are integrated with Artificial's orchestration platform to facilitate advanced molecular screening. It specifically targets the SARS-CoV-2 virus, focusing on the main protease, a critical enzyme in viral replication and a well-established drug target in Covid-19 therapy. This use case also showcases how this process can be automated, in other words, self-driving until certain criteria are met. This is potentially useful for the concept of self-driving labs where Artificial can orchestrate the workflows and integrate the AI models in the process. 

Based on the NVIDIA BioNeMo Blueprint for generative virtual screening, the self-driving virtual screening process is structured in iterative feedback loops, combining molecule selection, protein folding (if needed), docking, and binding affinity scoring. The workflow is executed as follows:

\begin{figure*}[ht]
    \centering
    \includegraphics[width=0.95\textwidth]{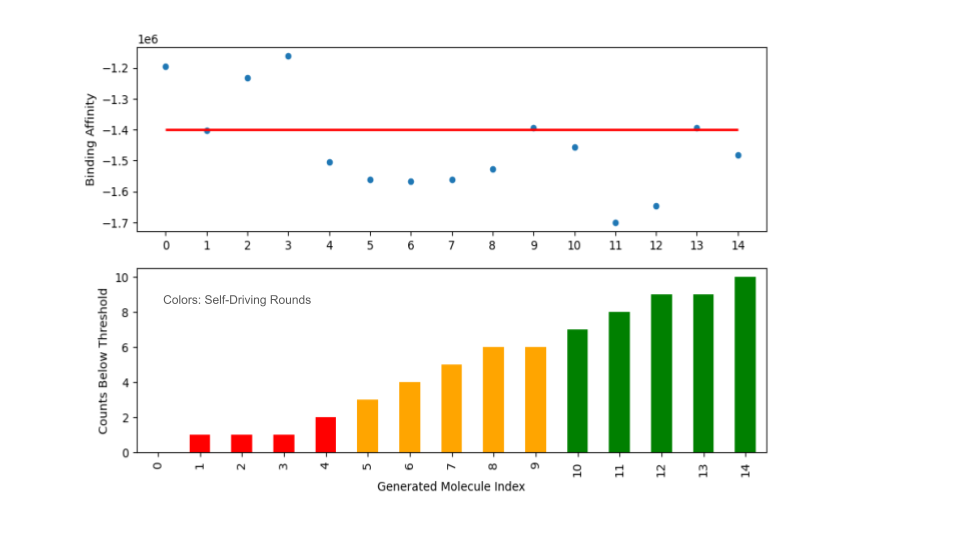}  % Reduce width slightly
    \caption{The self-driving iterations during the virtual screening proof of concept.}
    \label{fig:poc}
\end{figure*}

After three self-driving iterations, the set criteria were fulfilled:
\begin{enumerate}
\item The binding affinity threshold of -1.4 million was met.
\item A minimum of ten molecules were identified whose scoring values surpassed the required threshold.
\end{enumerate}
This iterative approach highlights the capability of Artificial's platform to autonomously refine molecular candidates, optimizing the selection process to focus on high-likelihood drug compounds. By combining BioNeMo's advanced molecular modeling tools with Artificial's automated orchestration and feedback systems, the PoC successfully demonstrated how AI-driven methodologies can accelerate virtual screening and drug discovery processes.

\section{Artificial’s infrastructure for model integration}

Artificial’s LabOps performs data management and integration for dry and wet labs, integrating purely in silico experimentation (as detailed here) with downstream laboratory automation. Central to Artificial’s architecture is the Lab Gateway, a network appliance that bridges the Artificial cloud to on-premises resources securely using bidirectional HTTP/2 streaming over an outbound connection. This allows LabOps to control and integrate data from automated laboratory equipment and secure IT infrastructure, including GPUs/TPUs for hosting AI models.

NVIDIA BioNeMo NIMs (containerized, pre-trained accelerated AI models, complete with APIs) allow models to be hosted in the cloud or securely behind a customer firewall on-premise. All data transacted between the Lab Gateway and any resource, cloud or on-premise, is kept in a permanent, immutable data record that is accessible via APIs or from other workflows. This allows data generated by inference and/or experimentation to be accessible for later training or fine-tuning of models.

\begin{figure*}[H]
    \centering
    \includegraphics[width=0.95\textwidth]{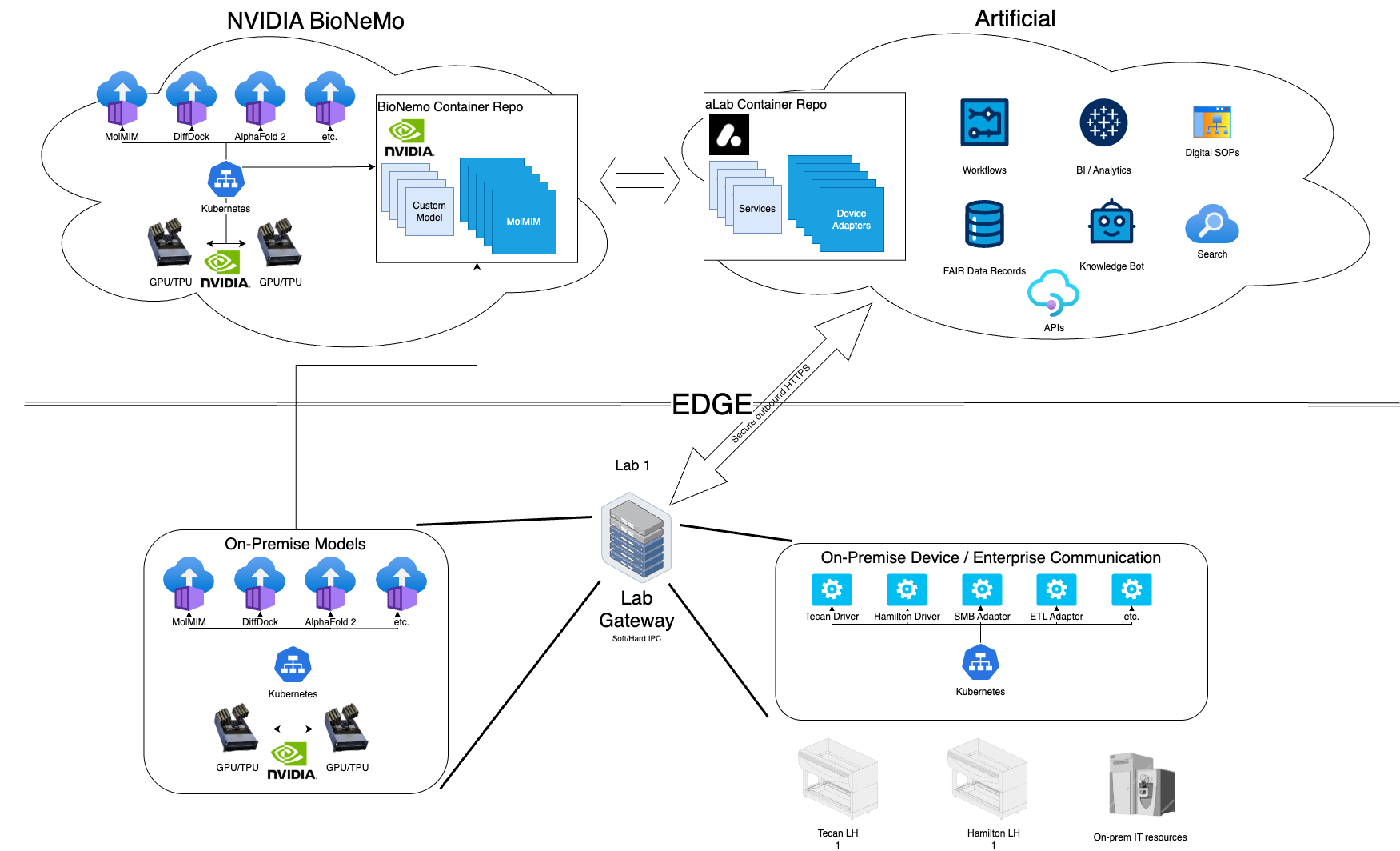}  % Reduce width slightly
    \caption{Artificial architecture for NVIDIA BioNeMo model integration.}
    \label{fig:poc}
\end{figure*}

Artificial’s ability to integrate AI models, such as those implemented in BioNeMo NIMs, into self-driving virtual screening workflows highlights its orchestration and data management capabilities. This approach extends beyond virtual screening, offering adaptability and scalability for a range of laboratory environments for drug discovery. By leveraging real-time simulation, sophisticated orchestration tools, and robust API integrations, Artificial’s platform enhances data quality, streamlines operations, and optimizes resource utilization, ultimately advancing both virtual screening and broader laboratory automation.

\section{Conclusions}
%%\label{}
In conclusion, Artificial serves as a transformative solution for self-driving labs, addressing critical challenges in orchestrating complex workflows, integrating diverse instruments and AI models, and managing data efficiently. By automating scheduling, decision-making, and real-time coordination of lab resources, Artificial enhances reproducibility, accelerates discovery, and optimizes lab operations. This system not only streamlines experimentation but also empowers researchers to leverage AI-driven insights for more efficient drug discovery. As self-driving labs continue to evolve, Artificial provides a scalable framework that will drive the next generation of breakthroughs in drug discovery and development.

\section*{Acknowledgments}
We would like to thank Janet Paulsen for helpful conversions regarding the BioNeMo proof of concept and Youssef Nashed for technical support on BioNeMo related questions.

\bibliographystyle{elsarticle-harv} 
\bibliography{example}

\begin{thebibliography}{42}
\expandafter\ifx\csname natexlab\endcsname\relax\def\natexlab#1{#1}\fi
\providecommand{\url}[1]{\texttt{#1}}
\providecommand{\href}[2]{#2}
\providecommand{\path}[1]{#1}
\providecommand{\DOIprefix}{doi:}
\providecommand{\ArXivprefix}{arXiv:}
\providecommand{\URLprefix}{URL: }
\providecommand{\Pubmedprefix}{pmid:}
\providecommand{\doi}[1]{\href{http://dx.doi.org/#1}{\path{#1}}}
\providecommand{\Pubmed}[1]{\href{pmid:#1}{\path{#1}}}
\providecommand{\bibinfo}[2]{#2}
\ifx\xfnm\relax \def\xfnm[#1]{\unskip,\space#1}\fi
%Type = Article
\bibitem[{Abolhasani and Kumacheva(2023)}]{abolhasani2023rise}
\bibinfo{author}{Abolhasani, M.}, \bibinfo{author}{Kumacheva, E.}, \bibinfo{year}{2023}.
\newblock \bibinfo{title}{The rise of self-driving labs in chemical and materials sciences}.
\newblock \bibinfo{journal}{Nature Synthesis} \bibinfo{volume}{2}, \bibinfo{pages}{483--492}.
%Type = Article
\bibitem[{Adediwura et~al.(2024)Adediwura, Koirala, Do, Wang and Miao}]{adediwura2024understanding}
\bibinfo{author}{Adediwura, V.A.}, \bibinfo{author}{Koirala, K.}, \bibinfo{author}{Do, H.N.}, \bibinfo{author}{Wang, J.}, \bibinfo{author}{Miao, Y.}, \bibinfo{year}{2024}.
\newblock \bibinfo{title}{Understanding the impact of binding free energy and kinetics calculations in modern drug discovery}.
\newblock \bibinfo{journal}{Expert Opinion on Drug Discovery} \bibinfo{volume}{19}, \bibinfo{pages}{671--682}.
%Type = Article
\bibitem[{Adelusi et~al.(2022)Adelusi, Oyedele, Boyenle, Ogunlana, Adeyemi, Ukachi, Idris, Olaoba, Adedotun, Kolawole et~al.}]{adelusi2022molecular}
\bibinfo{author}{Adelusi, T.I.}, \bibinfo{author}{Oyedele, A.Q.K.}, \bibinfo{author}{Boyenle, I.D.}, \bibinfo{author}{Ogunlana, A.T.}, \bibinfo{author}{Adeyemi, R.O.}, \bibinfo{author}{Ukachi, C.D.}, \bibinfo{author}{Idris, M.O.}, \bibinfo{author}{Olaoba, O.T.}, \bibinfo{author}{Adedotun, I.O.}, \bibinfo{author}{Kolawole, O.E.}, et~al., \bibinfo{year}{2022}.
\newblock \bibinfo{title}{Molecular modeling in drug discovery}.
\newblock \bibinfo{journal}{Informatics in Medicine Unlocked} \bibinfo{volume}{29}, \bibinfo{pages}{100880}.
%Type = Article
\bibitem[{Alharbi et~al.(2022)Alharbi, Gadiya, Henderson, Zaliani, Delfin-Rossaro, Cambon-Thomsen, Kohler, Witt, Welter, Juty et~al.}]{alharbi2022selection}
\bibinfo{author}{Alharbi, E.}, \bibinfo{author}{Gadiya, Y.}, \bibinfo{author}{Henderson, D.}, \bibinfo{author}{Zaliani, A.}, \bibinfo{author}{Delfin-Rossaro, A.}, \bibinfo{author}{Cambon-Thomsen, A.}, \bibinfo{author}{Kohler, M.}, \bibinfo{author}{Witt, G.}, \bibinfo{author}{Welter, D.}, \bibinfo{author}{Juty, N.}, et~al., \bibinfo{year}{2022}.
\newblock \bibinfo{title}{Selection of data sets for fairification in drug discovery and development: Which, why, and how?}
\newblock \bibinfo{journal}{Drug discovery today} \bibinfo{volume}{27}, \bibinfo{pages}{2080--2085}.
%Type = Article
\bibitem[{Alharbi et~al.(2023)Alharbi, Skeva, Juty, Jay and Goble}]{alharbi2023fair}
\bibinfo{author}{Alharbi, E.}, \bibinfo{author}{Skeva, R.}, \bibinfo{author}{Juty, N.}, \bibinfo{author}{Jay, C.}, \bibinfo{author}{Goble, C.}, \bibinfo{year}{2023}.
\newblock \bibinfo{title}{A fair-decide framework for pharmaceutical r\&d: Fair data cost--benefit assessment}.
\newblock \bibinfo{journal}{Drug discovery today} \bibinfo{volume}{28}, \bibinfo{pages}{103510}.
%Type = Article
\bibitem[{Blanco-Gonzalez et~al.(2023)Blanco-Gonzalez, Cabezon, Seco-Gonzalez, Conde-Torres, Antelo-Riveiro, Pineiro and Garcia-Fandino}]{blanco2023role}
\bibinfo{author}{Blanco-Gonzalez, A.}, \bibinfo{author}{Cabezon, A.}, \bibinfo{author}{Seco-Gonzalez, A.}, \bibinfo{author}{Conde-Torres, D.}, \bibinfo{author}{Antelo-Riveiro, P.}, \bibinfo{author}{Pineiro, A.}, \bibinfo{author}{Garcia-Fandino, R.}, \bibinfo{year}{2023}.
\newblock \bibinfo{title}{The role of ai in drug discovery: challenges, opportunities, and strategies}.
\newblock \bibinfo{journal}{Pharmaceuticals} \bibinfo{volume}{16}, \bibinfo{pages}{891}.
%Type = Article
\bibitem[{Blatter et~al.(2022)Blatter, Witte, Nakas and Leichtle}]{blatter2022big}
\bibinfo{author}{Blatter, T.U.}, \bibinfo{author}{Witte, H.}, \bibinfo{author}{Nakas, C.T.}, \bibinfo{author}{Leichtle, A.B.}, \bibinfo{year}{2022}.
\newblock \bibinfo{title}{Big data in laboratory medicine—fair quality for ai?}
\newblock \bibinfo{journal}{Diagnostics} \bibinfo{volume}{12}, \bibinfo{pages}{1923}.
%Type = Article
\bibitem[{Casalini(2021)}]{casalini2021not}
\bibinfo{author}{Casalini, T.}, \bibinfo{year}{2021}.
\newblock \bibinfo{title}{Not only in silico drug discovery: Molecular modeling towards in silico drug delivery formulations}.
\newblock \bibinfo{journal}{Journal of Controlled Release} \bibinfo{volume}{332}, \bibinfo{pages}{390--417}.
%Type = Article
\bibitem[{Cheng et~al.(2020)Cheng, Ma, Uzzi and Loscalzo}]{cheng2020importance}
\bibinfo{author}{Cheng, F.}, \bibinfo{author}{Ma, Y.}, \bibinfo{author}{Uzzi, B.}, \bibinfo{author}{Loscalzo, J.}, \bibinfo{year}{2020}.
\newblock \bibinfo{title}{Importance of scientific collaboration in contemporary drug discovery and development: a detailed network analysis}.
\newblock \bibinfo{journal}{BMC biology} \bibinfo{volume}{18}, \bibinfo{pages}{1--9}.
%Type = Incollection
\bibitem[{Colliandre and Muller(2023)}]{colliandre2023bayesian}
\bibinfo{author}{Colliandre, L.}, \bibinfo{author}{Muller, C.}, \bibinfo{year}{2023}.
\newblock \bibinfo{title}{Bayesian optimization in drug discovery}, in: \bibinfo{booktitle}{High Performance Computing for Drug Discovery and Biomedicine}. \bibinfo{publisher}{Springer}, pp. \bibinfo{pages}{101--136}.
%Type = Article
\bibitem[{Delgado-Licona and Abolhasani(2023)}]{delgado2023research}
\bibinfo{author}{Delgado-Licona, F.}, \bibinfo{author}{Abolhasani, M.}, \bibinfo{year}{2023}.
\newblock \bibinfo{title}{Research acceleration in self-driving labs: Technological roadmap toward accelerated materials and molecular discovery}.
\newblock \bibinfo{journal}{Advanced Intelligent Systems} \bibinfo{volume}{5}, \bibinfo{pages}{2200331}.
%Type = Article
\bibitem[{Deng et~al.(2022)Deng, Yang, Ojima, Samaras and Wang}]{deng2022artificial}
\bibinfo{author}{Deng, J.}, \bibinfo{author}{Yang, Z.}, \bibinfo{author}{Ojima, I.}, \bibinfo{author}{Samaras, D.}, \bibinfo{author}{Wang, F.}, \bibinfo{year}{2022}.
\newblock \bibinfo{title}{Artificial intelligence in drug discovery: applications and techniques}.
\newblock \bibinfo{journal}{Briefings in Bioinformatics} \bibinfo{volume}{23}, \bibinfo{pages}{bbab430}.
%Type = Article
\bibitem[{Denton et~al.(2021)Denton, Molloy, Charleston, Lipset, Hirsch, Mulberg, Howard and Marsh}]{denton2021data}
\bibinfo{author}{Denton, N.}, \bibinfo{author}{Molloy, M.}, \bibinfo{author}{Charleston, S.}, \bibinfo{author}{Lipset, C.}, \bibinfo{author}{Hirsch, J.}, \bibinfo{author}{Mulberg, A.E.}, \bibinfo{author}{Howard, P.}, \bibinfo{author}{Marsh, E.D.}, \bibinfo{year}{2021}.
\newblock \bibinfo{title}{Data silos are undermining drug development and failing rare disease patients}.
\newblock \bibinfo{journal}{Orphanet Journal of Rare Diseases} \bibinfo{volume}{16}, \bibinfo{pages}{1--4}.
%Type = Article
\bibitem[{Dodero-Rojas et~al.(2024)Dodero-Rojas, Contessoto, Fehlis, Mayala and Onuchic}]{Dodero-Rojas2024.07.17.603864}
\bibinfo{author}{Dodero-Rojas, E.}, \bibinfo{author}{Contessoto, V.G.}, \bibinfo{author}{Fehlis, Y.}, \bibinfo{author}{Mayala, N.}, \bibinfo{author}{Onuchic, J.N.}, \bibinfo{year}{2024}.
\newblock \bibinfo{title}{Epigenetics is all you need: A transformer to decode chromatin structural compartments from the epigenome}.
\newblock \bibinfo{journal}{bioRxiv} \URLprefix \url{https://www.biorxiv.org/content/early/2024/07/19/2024.07.17.603864}, \DOIprefix\doi{10.1101/2024.07.17.603864}.
%Type = Article
\bibitem[{Edfeldt et~al.(2024)Edfeldt, Edwards, Engkvist, G{\"u}nther, Hartley, Hulcoop, Leach, Marsden, Menge, Misquitta et~al.}]{edfeldt2024data}
\bibinfo{author}{Edfeldt, K.}, \bibinfo{author}{Edwards, A.M.}, \bibinfo{author}{Engkvist, O.}, \bibinfo{author}{G{\"u}nther, J.}, \bibinfo{author}{Hartley, M.}, \bibinfo{author}{Hulcoop, D.G.}, \bibinfo{author}{Leach, A.R.}, \bibinfo{author}{Marsden, B.D.}, \bibinfo{author}{Menge, A.}, \bibinfo{author}{Misquitta, L.}, et~al., \bibinfo{year}{2024}.
\newblock \bibinfo{title}{A data science roadmap for open science organizations engaged in early-stage drug discovery}.
\newblock \bibinfo{journal}{Nature Communications} \bibinfo{volume}{15}, \bibinfo{pages}{5640}.
%Type = Article
\bibitem[{Gadiya et~al.(2023)Gadiya, Ioannidis, Henderson, Gribbon, Rocca-Serra, Satagopam, Sansone and Gu}]{gadiya2023fair}
\bibinfo{author}{Gadiya, Y.}, \bibinfo{author}{Ioannidis, V.}, \bibinfo{author}{Henderson, D.}, \bibinfo{author}{Gribbon, P.}, \bibinfo{author}{Rocca-Serra, P.}, \bibinfo{author}{Satagopam, V.}, \bibinfo{author}{Sansone, S.A.}, \bibinfo{author}{Gu, W.}, \bibinfo{year}{2023}.
\newblock \bibinfo{title}{Fair data management: what does it mean for drug discovery?}
\newblock \bibinfo{journal}{Frontiers in Drug Discovery} \bibinfo{volume}{3}, \bibinfo{pages}{1226727}.
%Type = Article
\bibitem[{Glicksberg et~al.(2019)Glicksberg, Li, Chen, Dudley and Chen}]{glicksberg2019leveraging}
\bibinfo{author}{Glicksberg, B.S.}, \bibinfo{author}{Li, L.}, \bibinfo{author}{Chen, R.}, \bibinfo{author}{Dudley, J.}, \bibinfo{author}{Chen, B.}, \bibinfo{year}{2019}.
\newblock \bibinfo{title}{Leveraging big data to transform drug discovery}.
\newblock \bibinfo{journal}{Bioinformatics and Drug Discovery} , \bibinfo{pages}{91--118}.
%Type = Article
\bibitem[{Guo et~al.(2024)Guo, Meng, Lin, Zhou, Li, Tian and Huang}]{guo2024artificial}
\bibinfo{author}{Guo, S.B.}, \bibinfo{author}{Meng, Y.}, \bibinfo{author}{Lin, L.}, \bibinfo{author}{Zhou, Z.Z.}, \bibinfo{author}{Li, H.L.}, \bibinfo{author}{Tian, X.P.}, \bibinfo{author}{Huang, W.J.}, \bibinfo{year}{2024}.
\newblock \bibinfo{title}{Artificial intelligence alphafold model for molecular biology and drug discovery: a machine-learning-driven informatics investigation}.
\newblock \bibinfo{journal}{Molecular Cancer} \bibinfo{volume}{23}, \bibinfo{pages}{223}.
%Type = Article
\bibitem[{John et~al.(2024)John, Lin, Binder, Greaves, Shah, John, Lange, Hsu, Illango, Ramanathan et~al.}]{john2024bionemo}
\bibinfo{author}{John, P.S.}, \bibinfo{author}{Lin, D.}, \bibinfo{author}{Binder, P.}, \bibinfo{author}{Greaves, M.}, \bibinfo{author}{Shah, V.}, \bibinfo{author}{John, J.S.}, \bibinfo{author}{Lange, A.}, \bibinfo{author}{Hsu, P.}, \bibinfo{author}{Illango, R.}, \bibinfo{author}{Ramanathan, A.}, et~al., \bibinfo{year}{2024}.
\newblock \bibinfo{title}{Bionemo framework: a modular, high-performance library for ai model development in drug discovery}.
\newblock \bibinfo{journal}{arXiv preprint arXiv:2411.10548} .
%Type = Article
\bibitem[{Kasturi et~al.(2014)Kasturi, Brown, Brown, Madhavan, Prabakar and Wally}]{kasturi2014interconnectivity}
\bibinfo{author}{Kasturi, J.}, \bibinfo{author}{Brown, A.P.}, \bibinfo{author}{Brown, P.}, \bibinfo{author}{Madhavan, S.}, \bibinfo{author}{Prabakar, L.}, \bibinfo{author}{Wally, J.L.}, \bibinfo{year}{2014}.
\newblock \bibinfo{title}{Interconnectivity of disparate nonclinical data silos for drug discovery and development}.
\newblock \bibinfo{journal}{Therapeutic Innovation \& Regulatory Science} \bibinfo{volume}{48}, \bibinfo{pages}{498--506}.
%Type = Article
\bibitem[{Khanna(2012)}]{khanna2012drug}
\bibinfo{author}{Khanna, I.}, \bibinfo{year}{2012}.
\newblock \bibinfo{title}{Drug discovery in pharmaceutical industry: productivity challenges and trends}.
\newblock \bibinfo{journal}{Drug discovery today} \bibinfo{volume}{17}, \bibinfo{pages}{1088--1102}.
%Type = Article
\bibitem[{Lavecchia and Di~Giovanni(2013)}]{lavecchia2013virtual}
\bibinfo{author}{Lavecchia, A.}, \bibinfo{author}{Di~Giovanni, C.}, \bibinfo{year}{2013}.
\newblock \bibinfo{title}{Virtual screening strategies in drug discovery: a critical review}.
\newblock \bibinfo{journal}{Current medicinal chemistry} \bibinfo{volume}{20}, \bibinfo{pages}{2839--2860}.
%Type = Article
\bibitem[{Mak et~al.(2024)Mak, Wong and Pichika}]{mak2024artificial}
\bibinfo{author}{Mak, K.K.}, \bibinfo{author}{Wong, Y.H.}, \bibinfo{author}{Pichika, M.R.}, \bibinfo{year}{2024}.
\newblock \bibinfo{title}{Artificial intelligence in drug discovery and development}.
\newblock \bibinfo{journal}{Drug discovery and evaluation: safety and pharmacokinetic assays} , \bibinfo{pages}{1461--1498}.
%Type = Article
\bibitem[{Marco et~al.(2024)Marco, Evertsson, Riley, Tyrchan and Rathi}]{marco2024augmenting}
\bibinfo{author}{Marco, G.}, \bibinfo{author}{Evertsson, E.}, \bibinfo{author}{Riley, D.J.}, \bibinfo{author}{Tyrchan, C.}, \bibinfo{author}{Rathi, P.C.}, \bibinfo{year}{2024}.
\newblock \bibinfo{title}{Augmenting dmta using predictive ai modelling at astrazeneca}.
\newblock \bibinfo{journal}{Drug discovery today} , \bibinfo{pages}{103945}.
%Type = Article
\bibitem[{Moffat et~al.(2017)Moffat, Vincent, Lee, Eder and Prunotto}]{moffat2017opportunities}
\bibinfo{author}{Moffat, J.G.}, \bibinfo{author}{Vincent, F.}, \bibinfo{author}{Lee, J.A.}, \bibinfo{author}{Eder, J.}, \bibinfo{author}{Prunotto, M.}, \bibinfo{year}{2017}.
\newblock \bibinfo{title}{Opportunities and challenges in phenotypic drug discovery: an industry perspective}.
\newblock \bibinfo{journal}{Nature reviews Drug discovery} \bibinfo{volume}{16}, \bibinfo{pages}{531--543}.
%Type = Article
\bibitem[{Murugan et~al.(2022)Murugan, Priya, Sastry and Markidis}]{murugan2022artificial}
\bibinfo{author}{Murugan, N.A.}, \bibinfo{author}{Priya, G.R.}, \bibinfo{author}{Sastry, G.N.}, \bibinfo{author}{Markidis, S.}, \bibinfo{year}{2022}.
\newblock \bibinfo{title}{Artificial intelligence in virtual screening: Models versus experiments}.
\newblock \bibinfo{journal}{Drug Discovery Today} \bibinfo{volume}{27}, \bibinfo{pages}{1913--1923}.
%Type = Article
\bibitem[{Obaido et~al.(2024)Obaido, Mienye, Egbelowo, Emmanuel, Ogunleye, Ogbuokiri, Mienye and Aruleba}]{obaido2024supervised}
\bibinfo{author}{Obaido, G.}, \bibinfo{author}{Mienye, I.D.}, \bibinfo{author}{Egbelowo, O.F.}, \bibinfo{author}{Emmanuel, I.D.}, \bibinfo{author}{Ogunleye, A.}, \bibinfo{author}{Ogbuokiri, B.}, \bibinfo{author}{Mienye, P.}, \bibinfo{author}{Aruleba, K.}, \bibinfo{year}{2024}.
\newblock \bibinfo{title}{Supervised machine learning in drug discovery and development: Algorithms, applications, challenges, and prospects}.
\newblock \bibinfo{journal}{Machine Learning with Applications} \bibinfo{volume}{17}, \bibinfo{pages}{100576}.
%Type = Article
\bibitem[{Patel(2019)}]{patel2019overcoming}
\bibinfo{author}{Patel, J.}, \bibinfo{year}{2019}.
\newblock \bibinfo{title}{Overcoming data silos through big data integration}.
\newblock \bibinfo{journal}{International Journal of Computer Science and Technology} \bibinfo{volume}{3}.
%Type = Article
\bibitem[{Peeva et~al.(2025)Peeva, Guttman-Yassky, Yamaguchi, Berman, Oemar, Ramakrishna, Fasano, Evans-Molina, Chu, Ungar et~al.}]{peeva2025unlocking}
\bibinfo{author}{Peeva, E.}, \bibinfo{author}{Guttman-Yassky, E.}, \bibinfo{author}{Yamaguchi, Y.}, \bibinfo{author}{Berman, B.}, \bibinfo{author}{Oemar, B.}, \bibinfo{author}{Ramakrishna, J.}, \bibinfo{author}{Fasano, A.}, \bibinfo{author}{Evans-Molina, C.}, \bibinfo{author}{Chu, M.}, \bibinfo{author}{Ungar, B.}, et~al., \bibinfo{year}{2025}.
\newblock \bibinfo{title}{Unlocking disease insights to facilitate drug development: Pharmaceutical industry--academia collaborations in inflammation and immunology}.
\newblock \bibinfo{journal}{Drug Discovery Today} , \bibinfo{pages}{104317}.
%Type = Article
\bibitem[{Philippidis(2024)}]{philippidis2024gtc}
\bibinfo{author}{Philippidis, A.}, \bibinfo{year}{2024}.
\newblock \bibinfo{title}{Gtc 2024: Nvidia highlights ai ‘revolution’in drug discovery, genomics}.
\newblock \bibinfo{journal}{GEN Edge} \bibinfo{volume}{6}, \bibinfo{pages}{244--248}.
%Type = Article
\bibitem[{Pyzer-Knapp(2018)}]{pyzer2018bayesian}
\bibinfo{author}{Pyzer-Knapp, E.O.}, \bibinfo{year}{2018}.
\newblock \bibinfo{title}{Bayesian optimization for accelerated drug discovery}.
\newblock \bibinfo{journal}{IBM Journal of Research and Development} \bibinfo{volume}{62}, \bibinfo{pages}{2--1}.
%Type = Article
\bibitem[{Rapp et~al.(2024)Rapp, Bremer and Romero}]{rapp2024self}
\bibinfo{author}{Rapp, J.T.}, \bibinfo{author}{Bremer, B.J.}, \bibinfo{author}{Romero, P.A.}, \bibinfo{year}{2024}.
\newblock \bibinfo{title}{Self-driving laboratories to autonomously navigate the protein fitness landscape}.
\newblock \bibinfo{journal}{Nature chemical engineering} \bibinfo{volume}{1}, \bibinfo{pages}{97--107}.
%Type = Article
\bibitem[{Sanders et~al.(2023)Sanders, Scott, Yang, Qutub, Garcia~Martin, Berrios, Hastings, Rask, Mackintosh, Hoarfrost et~al.}]{sanders2023biological}
\bibinfo{author}{Sanders, L.M.}, \bibinfo{author}{Scott, R.T.}, \bibinfo{author}{Yang, J.H.}, \bibinfo{author}{Qutub, A.A.}, \bibinfo{author}{Garcia~Martin, H.}, \bibinfo{author}{Berrios, D.C.}, \bibinfo{author}{Hastings, J.J.}, \bibinfo{author}{Rask, J.}, \bibinfo{author}{Mackintosh, G.}, \bibinfo{author}{Hoarfrost, A.L.}, et~al., \bibinfo{year}{2023}.
\newblock \bibinfo{title}{Biological research and self-driving labs in deep space supported by artificial intelligence}.
\newblock \bibinfo{journal}{Nature Machine Intelligence} \bibinfo{volume}{5}, \bibinfo{pages}{208--219}.
%Type = Article
\bibitem[{Schauperl and Denny(2022)}]{schauperl2022ai}
\bibinfo{author}{Schauperl, M.}, \bibinfo{author}{Denny, R.A.}, \bibinfo{year}{2022}.
\newblock \bibinfo{title}{Ai-based protein structure prediction in drug discovery: impacts and challenges}.
\newblock \bibinfo{journal}{Journal of Chemical Information and Modeling} \bibinfo{volume}{62}, \bibinfo{pages}{3142--3156}.
%Type = Article
\bibitem[{Shukla and Tripathi(2021)}]{shukla2021molecular}
\bibinfo{author}{Shukla, R.}, \bibinfo{author}{Tripathi, T.}, \bibinfo{year}{2021}.
\newblock \bibinfo{title}{Molecular dynamics simulation in drug discovery: opportunities and challenges}.
\newblock \bibinfo{journal}{Innovations and implementations of computer aided drug discovery strategies in rational drug design} , \bibinfo{pages}{295--316}.
%Type = Article
\bibitem[{Udegbe et~al.(2024)Udegbe, Ebulue, Ebulue and Ekesiobi}]{udegbe2024machine}
\bibinfo{author}{Udegbe, F.C.}, \bibinfo{author}{Ebulue, O.R.}, \bibinfo{author}{Ebulue, C.C.}, \bibinfo{author}{Ekesiobi, C.S.}, \bibinfo{year}{2024}.
\newblock \bibinfo{title}{Machine learning in drug discovery: A critical review of applications and challenges}.
\newblock \bibinfo{journal}{Computer Science \& IT Research Journal} \bibinfo{volume}{5}, \bibinfo{pages}{892--902}.
%Type = Article
\bibitem[{Vergetis et~al.(2021)Vergetis, Skaltsas, Gorgoulis and Tsirigos}]{vergetis2021assessing}
\bibinfo{author}{Vergetis, V.}, \bibinfo{author}{Skaltsas, D.}, \bibinfo{author}{Gorgoulis, V.G.}, \bibinfo{author}{Tsirigos, A.}, \bibinfo{year}{2021}.
\newblock \bibinfo{title}{Assessing drug development risk using big data and machine learning}.
\newblock \bibinfo{journal}{Cancer research} \bibinfo{volume}{81}, \bibinfo{pages}{816--819}.
%Type = Inproceedings
\bibitem[{Wibowo et~al.(2017)Wibowo, Sulaiman and Shamsuddin}]{wibowo2017machine}
\bibinfo{author}{Wibowo, M.}, \bibinfo{author}{Sulaiman, S.}, \bibinfo{author}{Shamsuddin, S.M.}, \bibinfo{year}{2017}.
\newblock \bibinfo{title}{Machine learning in data lake for combining data silos}, in: \bibinfo{booktitle}{Data Mining and Big Data: Second International Conference, DMBD 2017, Fukuoka, Japan, July 27--August 1, 2017, Proceedings 2}, \bibinfo{organization}{Springer}. pp. \bibinfo{pages}{294--306}.
%Type = Article
\bibitem[{Zhou et~al.(2024)Zhou, Rusnac, Park, Canzani, Nguyen, Stewart, Bush, Nguyen, Wulff, Yarov-Yarovoy et~al.}]{zhou2024artificial}
\bibinfo{author}{Zhou, G.}, \bibinfo{author}{Rusnac, D.V.}, \bibinfo{author}{Park, H.}, \bibinfo{author}{Canzani, D.}, \bibinfo{author}{Nguyen, H.M.}, \bibinfo{author}{Stewart, L.}, \bibinfo{author}{Bush, M.F.}, \bibinfo{author}{Nguyen, P.T.}, \bibinfo{author}{Wulff, H.}, \bibinfo{author}{Yarov-Yarovoy, V.}, et~al., \bibinfo{year}{2024}.
\newblock \bibinfo{title}{An artificial intelligence accelerated virtual screening platform for drug discovery}.
\newblock \bibinfo{journal}{Nature Communications} \bibinfo{volume}{15}, \bibinfo{pages}{7761}.
%Type = Article
\bibitem[{Zhu(2020)}]{zhu2020big}
\bibinfo{author}{Zhu, H.}, \bibinfo{year}{2020}.
\newblock \bibinfo{title}{Big data and artificial intelligence modeling for drug discovery}.
\newblock \bibinfo{journal}{Annual review of pharmacology and toxicology} \bibinfo{volume}{60}, \bibinfo{pages}{573--589}.
%Type = Article
\bibitem[{Zhu et~al.(2022)Zhu, Zhang, Li and Huang}]{zhu2022comprehensive}
\bibinfo{author}{Zhu, H.}, \bibinfo{author}{Zhang, Y.}, \bibinfo{author}{Li, W.}, \bibinfo{author}{Huang, N.}, \bibinfo{year}{2022}.
\newblock \bibinfo{title}{A comprehensive survey of prospective structure-based virtual screening for early drug discovery in the past fifteen years}.
\newblock \bibinfo{journal}{International Journal of Molecular Sciences} \bibinfo{volume}{23}, \bibinfo{pages}{15961}.
%Type = Inproceedings
\bibitem[{Ziegler et~al.(2021)Ziegler, Reimann, Keller and Mitschang}]{ziegler2021metadata}
\bibinfo{author}{Ziegler, J.}, \bibinfo{author}{Reimann, P.}, \bibinfo{author}{Keller, F.}, \bibinfo{author}{Mitschang, B.}, \bibinfo{year}{2021}.
\newblock \bibinfo{title}{A metadata model to connect isolated data silos and activities of the cae domain}, in: \bibinfo{booktitle}{International Conference on Advanced Information Systems Engineering}, \bibinfo{organization}{Springer}. pp. \bibinfo{pages}{213--228}.

\end{thebibliography}



%% For example:

%% \bibitem[Aladro et al.(2015)]{Aladro15} Aladro, R., Martín, S., Riquelme, D., et al. 2015, \aas, 579, A101

%%\end{thebibliography}

\end{document}